\documentstyle[aps,manuscript]{revtex}
\begin{document}

\title{Time Dependent Effects and Transport Evidence for Phase Separation in
La$_{0.5}$Ca$_{0.5}$MnO$_{3}$}

\author{M. Roy$^{1}$, J. F. Mitchell$^{2}$, 
P. Schiffer$^{1}$\renewcommand{\thefootnote}{\alph{footnote}}
\footnote{schiffer.1@nd.edu}}

\address{$^{1}$Department of Physics, University of Notre Dame, IN 46556}
 
\address{$^{2}$Material Science Division, Argonne National Laboratory,
 Argonne, IL 60439} 


\maketitle
\begin{abstract}

{\hspace{0.5cm} The ground state of La$_{1-x}$Ca$_{x}$MnO$_{3}$
changes from a ferromagnetic metallic to an antiferromagnetic
charge-ordered state as a function of Ca concentration at x $\sim$ 0.50. We
present evidence from transport measurements on a sample with x = 0.50 that
the two phases can coexist, in agreement with other observations of phase
separation in these materials.  We also observe that, by applying and then
removing a magnetic field to the mainly charge-ordered state at some
temperatures, we can $``$magnetically anneal$"$ the charge order, resulting in a
higher zero-field resistivity.  We also observe logarithmic time dependence
in both resistivity and magnetization after a field sweep at low temperatures. }  

\end {abstract}

\newpage
\newpage
Among all the colossal magnetoresistance manganites
\cite{Jonker:50,Wollan:55,Searle:70,Matsu:70} 
La$_{1-x}$Ca$_{x}$MnO$_{3}$ is particularly interesting. It can be
prepared over the whole range of doping (x) (0 $\lesssim$ x $\lesssim$
1), and thus provides an insight into the properties of rare-earth
manganites as a function of doping. The ground state of this compound
ranges from a ferromagnetic metallic (FMM) at low doping (0.2 $\lesssim$
x $\lesssim$ 0.45) to an antiferromagnetic (AFM) charge-ordered (CO)
state at high doping x $\gtrsim$ 0.50. The concentration regime at 
x $\sim$ 0.50 \cite{ps:95,apr:96,chen:96,radaelli:97} is particularly 
interesting since in this regime the FM metallic state becomes 
unstable to an insulating charge-ordered state. However, recent 
magnetometry and neutron studies have detected a weak FM moment
of a few fraction of Bohr magneton/ion even in the charge-ordered 
regime. This can be explained by a homogeneous model, e.g., a 
canted AFM state or by a phase-separation into inhomogeneous 
state, although recent theoretical and experimental studies on 
this characteristic have shown overwhelming evidence for 
phase separation in this and other doped manganites 
\cite{ps:95,papavass:97,gong:95,radaelli:95,allodi:97,calvani:98,allodi:98,papavass:99,yunoki:98,moreo:99,schl:99,jaim:99,ueha:99,roy:98}.

In this paper, we report a study of La$_{0.5}$Ca$_{0.5}$MnO$_{3}$, 
where the ground state changes from a conducting FM (x $<$ 0.50) to a 
charge-ordered AFM state (x $\gtrsim$ 0.50) (additional data are also
presented in other publications \cite{roy:98,roy:99}).
The resistivity ($\rho$) was measured by standard ac four probe 
in-line method with the magnetic field perpendicular to the direction 
of the current, magnetization (M) with a commercially
available SQUID magnetometer, and heat capacity (C) was measured by
semi-adiabatic heat-pulse method. The sample was prepared by a standard solid
state reaction and the Mn$^{4+}$ content was measured to be 53.8$\%$  
by redox titration.

Figure 1 shows $\rho(T)$ and M(T) at different fields measured 
on both warming and cooling. On cooling, M(T) displays a 
FM transition at T$_{c}$, however, at a lower temperature 
T$_{co}$, M(T) drops sharply with an accompanying rise in 
$\rho(T)$ due to the charge-ordering of the Mn$^{4+}$ and Mn$^{3+}$ 
ions. Both M(T) and $\rho(T)$ exhibit large hysteresis at 
this transition, indicating the strongly first order nature of 
the transition. Furthermore T$_{co}$ decreases monotonically 
with increasing external field, indicating that the CO state 
becomes energetically less favorable in an external field \cite{roy:98}. 
The resistivity, $\rho(T)$, shows only activated behavior in 
low fields, but for H $\gtrsim$ 3 T, at temperatures well 
below T$_{co}$, $\rho(T)$ reaches a maximum before 
subsequently dropping at lower temperatures (see inset to \ref{f1}). 
The temperature of the maximum in $\rho(T)$ increases with 
increasing field, leading to an enormous magnetoresistance 
\cite{gong:95} which has been referred to as the $``$melting$"$ of 
the charge ordered state. The existence of a maximum and subsequent 
decrease in $\rho(T)$ can be attributed to the presence of free
carriers in the charge ordered state, and the large low temperature
magnetoresistance can be attributed to an increase in the population of the
free carriers.   The solid line in the 
inset is a fit to the data assuming the coexistence of free 
carriers and charge order as has been described previously 
\cite{roy:98}. Fits to the $\rho(T)$ data at different 
fields indicate that the $``$melting$"$ proceeds by an 
increase in the number of free carriers, but that even in 
a 9 T field only a small fraction of the charge 
order is dissociated, as is also indicated by a rather 
large $\rho$ ($\rho$ $\sim$ 0.1 $\Omega$cm) at high 
fields.  It should be noted that this sort of coexistence is seen in a
range of samples with 0.48$\lesssim$ x $\lesssim$ 0.55 in 
La$_{1-x}$Ca$_{x}$MnO$_{3}$ and also in other materials 
including high quality single crystals \cite{toku:96}.

To characterize the dissociation of the CO state in an external magnetic
field, we studied $\rho(H)$ and M(H) as a function of field as shown in
figure \ref{f2}. Our samples were zero-field cooled to the prescribed 
temperature and $\rho(H)$ and M(H) were measured as a function of 
field, when the field was swept from H = 0 $\rightarrow$ H$_{max}$ , 
H$_{max}$  $\rightarrow$ -H$_{max}$  and -H$_{max}$  $\rightarrow$ H$_{max}$, 
where H$_{max}$ was 9 T for the resistivity and 7 T for the
magnetization measurements. At low temperatures (T $\lesssim$ 60), 
$\rho(H)$ drops with increasing field \cite{xiao:96}. During subsequent
field sweeps, while $\rho(H)$ displays a large hysteresis, M(H) 
remains largely non-hysteretic at high fields. On decreasing the 
field $\rho(H)$ increases, but it always remains considerably smaller 
than during the initial sweep. This 
perhaps suggests that even though the charge-lattice is not 
totally dissociated at H $\sim$ 9 T, at the lowest temperatures 
the delocalized electrons remain primarily dissociated even when the field is 
removed. At intermediate temperatures (70 $\gtrsim$ T $\gtrsim$ 140), where the 
conduction is primarily through excitations in the charge lattice,
$\rho(H)$ also decreases with increasing field. At lower fields 
(H $\lesssim$ 1 T), however, $\rho(H)$ rises above the
initial sweep, such that $\rho(H = 0)$ is higher than the initial
ZFC $\rho$ of the sample. Similar behaviour was also observed in high
quality single crystal samples of Pr$_{0.50}$Sr$_{0.50}$MnO$_{3}$
\cite{tomi:95} and Nd$_{0.50}$Sr$_{0.50}$MnO$_{3}$ \cite{roy:99}. 
We speculate this increase in $\rho(H)$ at low fields is due to 
field-induced $``$annealing$"$ of the charge-ordered state, 
i.e., by sweeping the field up and back, more perfect charge-ordered
states with correspondingly higher resistivity are created.  At temperatures
of the order of T$_{co}$, this annealing effect disappears but the sample
continues to show large magnetoresistance. This is probably also
attributable to the enhancement of the ferromagnetic phase in the 
sample, but at these temperatures the enhancement is apparently 
reversible since the $\rho(H=0)$ is recovered after sweeping the field.

The competition between the phase separated AFM CO and FMM states
can be expected to lead to interesting time dependent effects of 
the sort seen in spin glasses (due to the local frustration between 
AFM and FM exchanges \cite{mydo:93}). In particular, since the relative 
fraction of the phase separated materials can be altered by the 
application of an external magnetic field, one might expect to 
observe time dependence in the physical properties which is 
attributable to the changing phase separation. In order to examine 
such effects, we measured the resistivity, magnetization and 
heat capacity of this sample as function of time after changing 
the magnetic field. In each case, the sample is zero field 
cooled to the prescribed temperature, the field was raised, and 
then measurements were performed as a function of time. Both 
zero field cooled $\rho$ and M relax monotonically as function
of time in an external field with a logarithmic time dependence 
as shown in figure \ref{f3}, but C which shows a sharp rise in 
the first 10 minutes after applying a 9 T field remains 
time-independent afterwards. Furthermore, the time dependence might 
be more pronounced in transport preoperties than the magnetization
since one can imagine that there are parallel conductive paths, 
which would increase the conductivity of the sample. And we
observe that while the magnetization increases by only 
around 0.5$\%$ over a period of 24 hours, $\rho$ decreases
by as much as 60$\%$ in the same period of time.   
The time dependence of M and $\rho$ is similar to  
magnetic viscosity or magnetic after-effects observed in 
soft irons \cite{stre:49} and spin glasses. Similar 
time-dependent relaxation in $\rho$ was also 
observed in La$_{2/3}$Ca$_{1/3}$MnO$_{3+\delta}$ \cite{helm:95}, 
and Pr$_{2/3}$Ca$_{1/3}$MnO$_{3}$ \cite{anan:99} which were also
observed to display evidence of some glassy behaviour by 
recent $\mu$ spin resonance \cite{heff:96} and neutron 
diffraction \cite{yosh:95} experiments. Figure \ref{f3}
shows that at H $=$ 7 T and 9 T respectively, both M and $\rho$ 
can be fitted to a linear function of logarithm of time, i.e., 
$M = Slogt+const.$. A preliminary study of the slope (S) as
a function of temperature reveals some interesting features,
and which are the subject of ongoing investigation.


This research has been supported by NSF grant DMR 97-01548,
the Alfred P. Sloan Foundation and the Dept. of Energy, Basic Energy
Sciences-Materials Sciences under contract $\#$W-31-109-ENG-38.


\begin{figure}

\caption{\label {f1} The top panel shows the resistivity of the sample as a 
function of temperature on cooling(solid) and warming(open) at 
H $=$ 0 $\rightarrow$ 9 T in steps of 1 T. The bottom panel shows the
magnetization of the sample as a function of temperature on 
cooling(solid) and warming(open) at H $=$ 1 T $\rightarrow$ 7 T 
in steps of 2 T. The inset illustrates two distinct features in
$\rho(T)$ associated with the coexistence of two states. The solid
line is fit to the data as discussed in the text.} 

\end {figure}

\begin{figure}

\caption{\label {f2} The left and the right panels show the resistivity and
magnetization of the sample as a function of field at different 
temperatures. All the measurements were done when the field was swept from 0 
$\rightarrow$ 9 T (open circles), 9 T $\rightarrow$ -9 T 
(solid lines) and -9 T $\rightarrow$ 9 T 
(dashed line).}

\end{figure}

\begin{figure}

\caption{\label {f3} The time dependence of $\rho(t)$ and M(t) at T $=$ 5 K and
H = 9 T and 7 T respectively. Both resistivity and magnetization data is fitted 
with the functional form, $\rho = Slnt+const.$ (solid line). In addition a fit of the form $\rho = Aexp(-BT)$ is also illustrated by dashed lines. The inset shows 
C(t) at H $=$ 9 T and T $=$ 5.5 K.}
\end{figure}

\end{document}